\documentclass[journal = jacs, manuscript=article]{achemso}
\usepackage{graphicx}
\usepackage{color}
\usepackage{caption}
\usepackage{amssymb}
\usepackage{amsmath}
\usepackage{gensymb}
\usepackage{booktabs}

\author{Mohammad R. Momeni$^{*}$}
\author{Zeyu Zhang} 
\author{David Dell'Angelo}
\author{Farnaz A. Shakib}
\email{momeni@njit.edu, shakib@njit.edu}
\affiliation{Department of Chemistry and Environmental Science, New Jersey Institute of Technology, Newark 07102, NJ United States}

\title{Unravelling Water Stability and Electrical Conductivity of 2D Layered Metal-Organic Frameworks in Aqueous Solutions}

\begin{document}

\begin{abstract}
Molecular dynamics simulations combined with periodic electronic structure calculations are performed to decipher structural, thermodynamical and dynamical properties of the interfaced vs. confined water adsorbed in hexagonal 1D channels of the 2D layered electrically conductive Cu$_3$(HHTP)$_2$ and Cu$_3$(HTTP)$_2$ metal-organic frameworks (HHTP$=$2,3,6,7,10,11-hexahydroxytriphenylene and HTTP = 2,3,6,7,10,11-hexathiotriphenylene). Comparing water adsorption in bulk vs. slab models of the studied 2D MOFs shows that water is preferentially adsorbed on the framework walls via forming hydrogen bonds to the organic linkers rather than by coordinating to the coordinatively unsaturated open-Cu$^{2+}$ sites. Theory predicts that in Cu$_3$(HTTP)$_2$ the van der Waals interactions are stronger which helps the MOF maintain its layered morphology with allowing very little water molecules to diffuse into the interlayer space. Data presented in this work are general and helpful in implementing new strategies for preserving the integrity as well as electrical conductivity of porous materials in aqueous solutions.
\end{abstract}

\clearpage

\section{Introduction}
\begin{figure}[!h]
\centering
  \includegraphics[scale=0.52]{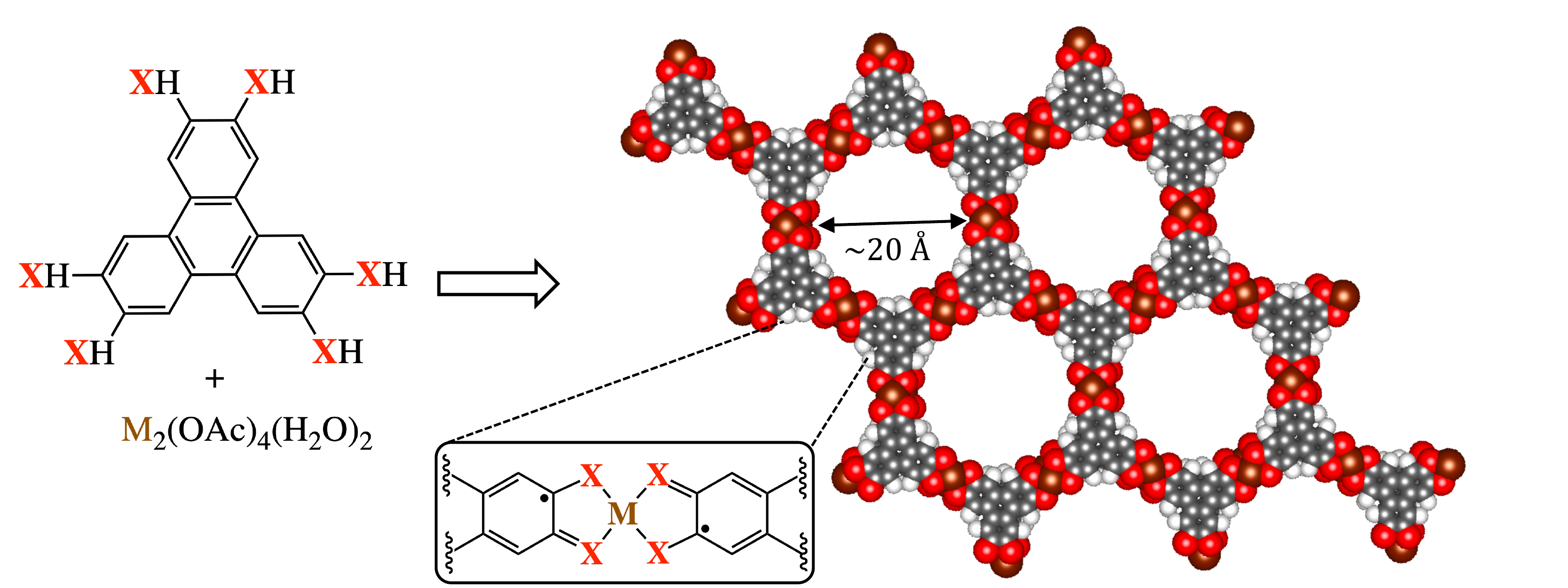}
  \caption{{\color{black}{Structure of 2D MOFs. Tritopic linkers and the self-assembled layered architecture along the main 1D channel with M$=$Cu$^{2+}$ and X$=$O (Cu$_3$(HHTP)$_2$) and S (Cu$_3$(HTTP)$_2$) in this work.}}}
  \label{Figure1}
\end{figure}

Conductive $\pi-$stacked 2D layered metal-organic frameworks (MOFs)\cite{Hmadeh:2012} are a new addition to the family of nanoporous materials which offer electrical conductivity \cite{Huang:2015,Sun:2016,Clough:2017,Dou:2017} in addition to permanent porosity and exceptionally high surface area of conventional 3D MOFs. Crystal structures of 2D MOFs are composed of tetra-coordinated semi-planar metal nodes connected through aromatic linkers creating extended $\pi-$conjugated 2D layers in \textit{ab} plane. Stacking of layers \textit{via} $\pi-\pi$ interactions between aromatic triphenylene rings creates parallel, noninterconnected, 1D infinite hexagonal channels along the \textit{c} axis. 
This architecture provides both \textit{in}-plane and \textit{out}-of-plane electrical conduction as well as directional permeation of electrolyte and target molecules through the conductive material, Figure~\ref{Figure1}. Owing to this ideal combination, unprecedented breakthroughs are made possible in producing cost-effective semiconductors;\cite{Sheberla:2014,Wu:2017} supercapacitors;\cite{Sheberla:2016,Li:2017,Sheberla:2017,Feng:2018} and ion-to-electron transduced chemical sensors.\cite{Clough:2015,Dong:2015,Miner:2016,Huang:2017,Sun:2017,Mendecki:2017,Yao:2017,Mendecki:2018,Jia:2018,Downes:2018,Meng:2019} {\color{black}{Progress towards these applications calls for a full investigation of stability of $\pi-$stacked layered MOFs subject to humidity, and possible hydrolysis of the MOF secondary building unit (SBU), and the effects of these structural transformations on the overall electrical conductivity of the material.}} 

When it comes to 2D MOFs, little, if not none, experimental data is available on the dynamics of confined water and its adsorption by the framework. There are many important questions to be answered; (i) What are the parameters governing the dynamics of confined water along the 1D channels versus penetration into the interlayer space; (ii) How do the van der Waals interactions reply to the presence of water within the interlayer space; (iii) Will layers be separated due to presence of water or the interlayer water will act as a glue to keep  layers together?; {\color{black}{(iv) Regardless, how will the presence of water in interlayer space affect the charge mobility between layers and the overall electrical conductivity of the 2D MOFs?}} and many other such questions. 

{\color{black}{Cu$_3$(HHTP)$_2$ (HHTP$=$2,3,6,7,10,11-hexahydroxytriphenylene) is one of the first conductive 2D layered MOFs with metallic behaviour that was synthesized alongside its Co and Ni counterparts in 2012.\cite{Hmadeh:2012} In our recent study on stability of Cu$_3$(HHTP)$_2$ and Co$_3$(HHTP)$_2$ in aqueous solutions,\cite{Shi:2020} we introduced organic linker as an effective player in the dynamics of confined water in 1D channels as well as interlayer space of 2D MOFs.}}
We expect that water is first adsorbed to the 1D channel walls \textit{via} hydrogen bond (HB) formation with organic linkers. The flexible nature of layers and the dynamic HB network finally allows penetration of water molecules within the interlayer space. The penetrated water molecules can then form coordinative bonds with open-metal sites which leads to the increase of interlayer distance and weakening of van der Waals interactions, possibly affecting the overall integrity and electrical conductivity of the layered structure. To verify this hypothesis, here we choose to study dynamics of interface vs. confined water in the slab and bulk models of Cu$_3$(HHTP)$_2$ versus Cu$_3$(HTTP)$_2$ {\color{black}{(HTTP$=$ 2,3,6,7,10,11-hexathiotriphenylene)}}, depicted in Figure~\ref{Figure1}, as representatives of $\pi-$stacked 2D layered MOFs. {\color{black}{We will follow the structural, thermodynamical, and dynamical footprints of water in 2D MOFs. Along the path, we will evaluate their effects on the electronic band structure of Cu$_3$(HHTP)$_2$ and Cu$_3$(HTTP)$_2$ and will differentiate the \textit{in}-plane and \textit{out}-of-plane charge mobility routes.}} These findings are transferable and applicable to other similar 2D layered nanoporous materials and are useful in designing more robust water stable materials with desired applications.

\section{Results and discussion}
\subsection{Dynamical picture of the interaction of interface vs. confined water with 2D MOFs}

\begin{figure*}[!h]
\centering
  \includegraphics[width=0.99\linewidth]{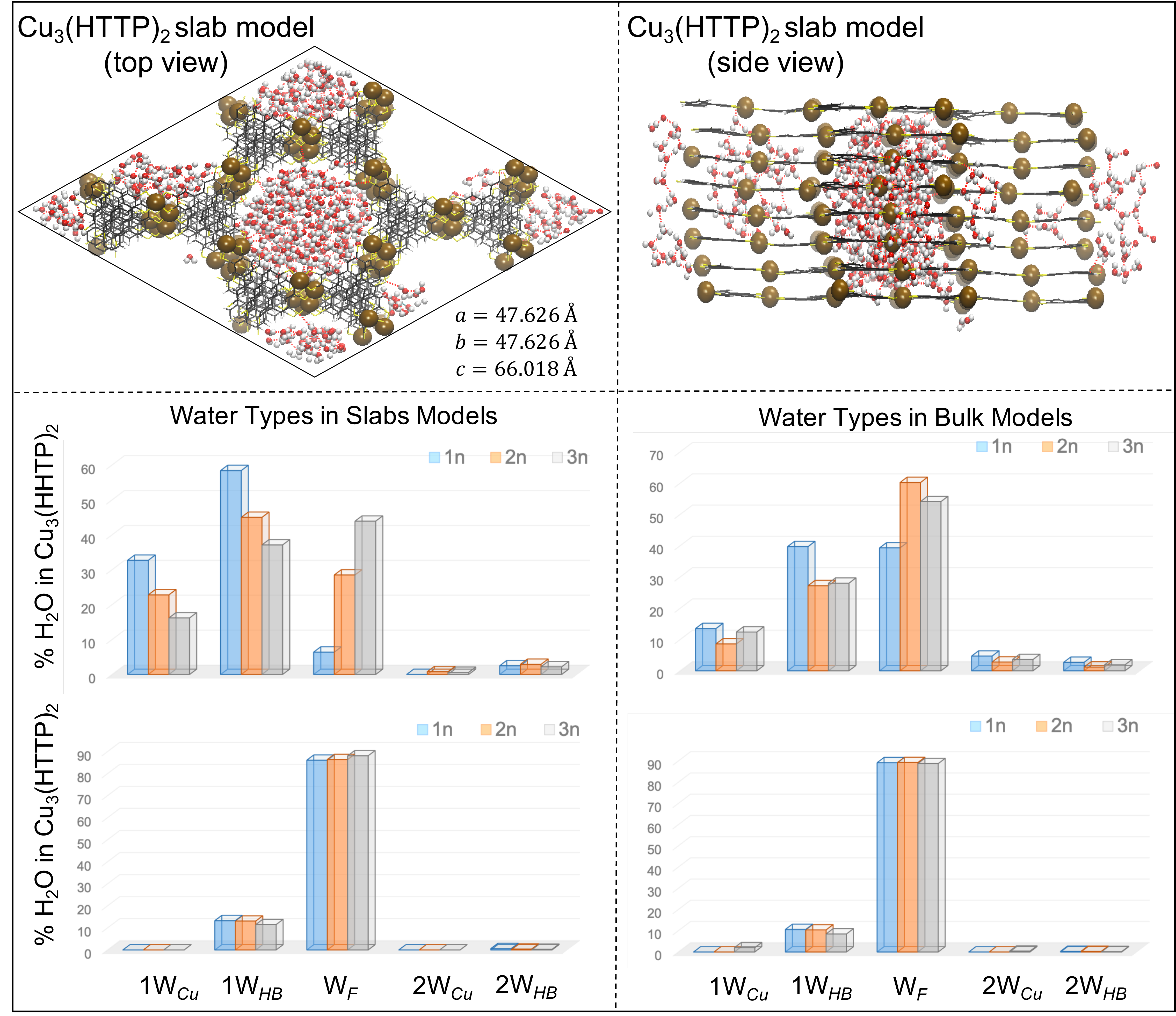}
  \caption{Top: top and side views of representative snapshots for equilibrated slab models of Cu$_3$(HTTP)$_2$ with the highest water loading of 384 H$_2$O {\color{black}{and with the equilibrated dimensions of the simulations box given.}} Calculated percentage different types of water molecules in the slab (left) and bulk (right) models of Cu$_3$(HHTP)$_2$ (middle) and Cu$_3$(HTTP)$_2$ (bottom) MOFs with different water loadings. \textit{n} refers to the number of Cu$^{2+}$ centers in the unit cell which is equal to 48 in the bulk and 96 in the slab models. {\color{black}{Water molecules coordinated to one or two open-Cu$^{2+}$ sites (1W$_{Cu}$ and 2W$_{Cu}$), hydrogen bonded (HB) to one or two oxygen or sulfur atoms of the linkers (1W$_{HB}$ and 2W$_{HB}$) as well as free water molecules (W$_F$) for both bulk and slab models are shown.}}}
  \label{Figure2}
\end{figure*}

{\color{black}{All hydrated systems, both bulk and slab models, were equilibrated in the isothermal-isobaric NPT ensemble for 5 ns. The equilibrated systems are provided in the SI Figure S6 while the final equilibrated slab model of Cu$_3$(HTTP)$_2$, with 384 water molecules, is given in Figure \ref{Figure2} as an example.}} Analysis of water layers in the equilibrated systems reveals five different types of water considering their interactions with the framework. Figure \ref{Figure2} depicts the percentage of water molecules coordinated to one or two open-Cu$^{2+}$ sites (1W$_{Cu}$ and 2W$_{Cu}$), water molecules hydrogen bonded (HB) to one or two oxygen or sulfur atoms of the linkers (1W$_{HB}$ and 2W$_{HB}$) and free water molecules (W$_F$) for both bulk and slab models.
Calculated smaller percentage of W$_F$s in slabs of Cu$_3$(HHTP)$_2$ compared to Cu$_3$(HTTP)$_2$, regardless of water concentration, clearly shows higher affinity of the former to adsorb water than the latter. Though the metal centers have the same identity in both frameworks but both 1W$_{Cu}$ and 2W$_{Cu}$ are negligible in Cu$_3$(HTTP)$_2$ slabs. Confirming our working hypothesis, HB formation with organic linkers are the first step of adsorption. As oxygen atoms of the HHTP linkers form rather stronger HBs with water molecules than sulfur of HTTP, Cu centers in Cu$_3$(HHTP)$_2$ are more exposed to the water molecules. Furthermore, the open-Cu$^{2+}$ centers connected to electronegative oxygen atoms in Cu$_3$(HHTP)$_2$ are more positively charged compared to the open-Cu$^{2+}$ centers connected to sulfur in Cu$_3$(HTTP)$_2$ (SI Tables S1 and S6). The other evidence for our hypothesis is the observed trend in 1W$_{Cu}$ with respect to water concentration that closely follows the trend of 1W$_{HB}$, Figure \ref{Figure2}. The bulk models, more or less, follow the same trends. To illustrate a more informative picture of dynamics of water, we calculated water reorientation relaxation times ($\tau_2$ in ps) and diffusion coefficients along the interlayer space (D$_{xy}$) and the main 1D channel (D$_z$) as well as total diffusion coefficients (D$_{tot}$), see Table~\ref{Table1}. We calculated D$_{tot}$ based on the mean square displacement of the particles \cite{Nitzan:2006} using the following equation:
\begin{equation}
D=\lim_{t \to \infty} \frac{1}{6t}\langle(\textbf{r}(t)-\textbf{r}(0))^2\rangle
\label{eqn1}
\end{equation}
We quantified water reorientation relaxation times in terms of the orientational time correlation functions (TCFs). Then, formulated TCFs in terms of the reorientation of the unit vector $\hat{\textbf{u}}_{OH}$ that lies along one of the OH bonds of a water molecule:
\begin{equation}
C_{2,\textrm{OH}}(t)= \langle P_2 [\hat{\textbf{u}}_{\textrm{OH}}(0)  \hat{\textbf{u}}_{\textrm{OH}}(t)] \rangle
\label{eqn2}
\end{equation} 
where $P_2$ is the Legendre polynomial of order $2$. The time dependence of the 2$^{nd}$ Legendre polynomial is exponential and can be used to determine orientational relaxation time ($\tau_2$):
\begin{equation}
C_{2,\textrm{OH}}(t) \propto \textrm{exp}(-\frac{t}{\tau_2}).
\label{eqn3}
\end{equation} 

\begin{table}[htbp]
\caption{\label{Table1} Calculated water reorientation relaxation time ($\tau_2$ in ps), diffusion coefficients along the \textit{xy} plane (D$_{xy}$) and \textit{z} direction (D$_z$) as well as total diffusion coefficients (D$_{tot}$) in \AA$^2\cdot$ps$^{-1}$ for bulk and slab (in parenthesis) models of 2D MOFs with different water loadings. Corresponding experimental values are provided for comparison.}
\begin{tabular}{ccccc}
\hline
\multicolumn{5}{c} {\textbf{Cu$_3$(HHTP)$_2$}}\\ 
\cline{2-5}
\textbf{n} & $\tau_2$ & D$_{tot}$ & D$_{xy}$  & D$_{z}$  \\
\hline 
\textbf{1H$_2$O/Cu$^{2+}$}& 18.8 (59.9) & 0.063 (0.030) & 0.050 (0.028) & 0.089 (0.036)\\
\textbf{2H$_2$O/Cu$^{2+}$}& 17.2 (47.1) & 0.051 (0.031) & 0.047 (0.029) & 0.059 (0.034) \\
\textbf{4H$_2$O/Cu$^{2+}$}& 25.4 (30.5) & 0.039 (0.036) &0.033 (0.035) &0.050 (0.038) \\
\hline
\multicolumn{5}{c} {\textbf{Cu$_3$(HTTP)$_2$ }}\\ 
\cline{2-5}
\textbf{n} & $\tau_2$ & D$_{tot}$ & D$_{xy}$  & D$_{z}$  \\
\hline 
\textbf{1H$_2$O/Cu$^{2+}$}& 3.1 (4.4) &0.225 (0.232) & 0.149 (0.249) & 0.376 (0.199)\\
\textbf{2H$_2$O/Cu$^{2+}$}& 4.7 (4.7) & 0.150 (0.165) & 0.111 (0.163)& 0.228 (0.167)\\
\textbf{4H$_2$O/Cu$^{2+}$}& 5.8 (5.3) & 0.106 (0.129)  & 0.095 (0.129) & 0.126 (0.129)\\
\hline
\textbf{Exp. bulk water}&1.7$-$2.6\cite{Winkler:2000,Lawrence:2003,Rezus:2005}&0.229\cite{Krynicki:1978}\\
\hline
\end{tabular}
\end{table}

\begin{figure*}
 \centering
 \includegraphics[width=0.99\linewidth]{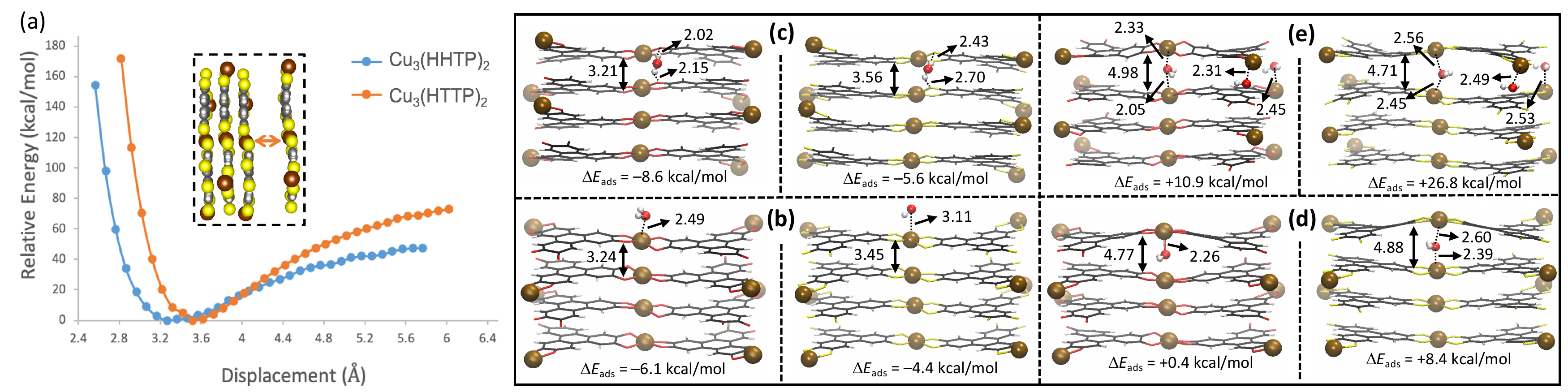}
 \caption{(a) Calculated relative energies (kcal/mol, PBE-D3) for dry slab models of the studied 2D MOFs with minima located at 3.267 Å and 3.519 Å, representing the interlayer distances. To obtain these graphs the top outermost layers are scanned starting from their respective PBE-D3 optimized geometries. (b)-(e) Key bond and interlayer distances (\AA, PBE-D3) of different modes of water interacting with the slab models of the Cu$_3$(HHTP)$_2$ (left) and Cu$_3$(HTTP)$_2$ (right) 2D MOFs.}
 \label{Figure3}
\end{figure*}
 Comparison of $\tau_2$ values between either bulk or slab models of Cu$_3$(HHTP)$_2$ and Cu$_3$(HTTP)$_2$ again shows the free nature of water in the latter with calculated $\tau_2$ values in the range of 3.1-5.8 ps close to that of bulk water, Table~\ref{Table1}. In comparison, a range of $\tau_2$ values of 17.2-59.9 ps was found for Cu$_3$(HHTP)$_2$. This is in line with the computed D$_{tot}$ values which are 3-8 times higher in Cu$_3$(HTTP)$_2$ than Cu$_3$(HHTP)$_2$, Table~\ref{Table1}. A more informative picture of water dynamics can be obtained from comparison between D$_{xy}$ and D$_{z}$ since they can shed light on water mobility into the interlayer space vs. the main 1D channel, respectively. In the case of bulk models, D$_{z}$ is always higher than D$_{xy}$ which shows water molecules cannot move as freely along the \textit{xy} plane as the \textit{z} direction. Focusing on bulk Cu$_3$(HHTP)$_2$, one can notice a decreasing trend of D$_{z}$ with increasing water content. This can refer to the adsorption of water molecules by the framework where coordinated water molecules to the metal centers act as nucleation sites and draw more water within the interlayer space, overall reducing their mobility alongside the 1D channel. However, there is also a decreasing trend in D$_{xy}$ with increasing water content. Analysis of MD trajectories, movies are provided as part of SI, shows that as the interlayer space is saturated with water, some are able to penetrate to the nearby channel but due to the high affinity of the framework for adsorbing water they still stay close to the surface. Calculated diffusion coefficients for bulk Cu$_3$(HTTP)$_2$ show rather similar trends to those of Cu$_3$(HHTP)$_2$ in spite of low affinity of the framework toward water (top panels of Figure~\ref{Figure2}). In this case, the driving force for the dynamics of water in \textit{xy} plane can be a network of hydrogen bonded water molecules that move freely between MOF layers and reaches the nearby channel. Analysis of MD trajectories confirms this hypothesis as the penetrated water molecules in the next channel now form a droplet away from the framework surface. Slab models do not behave as systematically as bulk models since water molecules experience two different environments in the \textit{xy} direction, one is the interlayer space similar to the bulk model, the other is the interface of the slab with water in the \textit{z} direction. One certain point that can be drawn from the data in Table~\ref{Table1} regarding slab models is that they generally show smaller D$_z$ values than their counterpart bulk models. The reason is that incoming water molecules toward slab first disperse on the interface where there is plenty of open-metal sites without steric hindrance of an adjacent layer. This overall hinders the ability of the water molecules to enter into the main 1D channels. A combined picture of the dynamics of water in the slab and bulk models shows that in a realistic compact device, incoming water, and any substrate within, would face a physical hindrance to enter the channels of Cu$_3$(HHTP)$_2$. After entering the channel, their movement would still be slow due to penetration into the interlayer space. A device made of Cu$_3$(HTTP)$_2$ on the other hand would create less physical hindrances for the movement of water, and any other solvated substrates, however, there might be disadvantages of reduced contact time between framework and substrate if performing a catalytic reaction is of interest. {\color{black}{In any case, penetration of water within interlayer space is evident regardless of nature of organic linkers. This is expected to lead to an increase in the interlayer distance and possibly weakening of the $\pi-\pi$ and d-$\pi$ van der Waals interactions. Now, the question is how these structural changes are going to affect the overall stability and electrical conductivity of the layered framework in aqueous solutions?}}

\subsection{Thermodynamical footprints of water on structural deformations of 2D MOFs}
Our MD simulations discussed in the previous section show that penetration of water molecules within interlayer space can occur regardless of the hydrophilicity or hydrophobicity of the framework driven by the osmosis effect of the vacant nearby channels. Will this phenomenon result in the separation of layers from each other or water molecules can act as a glue to keep layers together? We turn to quantum mechanical calculations to answer this and similar questions. First, we build a 1$\times$1$\times$2 slab model comprised of four layers for both Cu$_3$(HHTP)$_2$ and Cu$_3$(HTTP)$_2$ 2D MOFs suitable for such calculations. To gauge and compare the strength of $\pi-\pi$ interactions in the two systems in the absence of water, the top outermost layer was scanned from 2.567~\AA~-- 5.567~\AA~in Cu$_3$(HHTP)$_2$ and from 2.819~\AA~-- 6.019~\AA~in Cu$_3$(HTTP)$_2$, Figure \ref{Figure3}(a). The minimum of energy occurs at 3.267 \AA~for Cu$_3$(HHTP)$_2$ and 3.519 \AA~for Cu$_3$(HTTP)$_2$ representing the interlayer distance in optimized structures. The plateau represent dissociation energy of the outermost layer from the slab which is $\sim$20 kcal/mol higher for Cu$_3$(HTTP)$_2$ than Cu$_3$(HHTP)$_2$. This indicates stronger $\pi-\pi$ interactions in the former likely due to the larger atomic radius and polarizability of sulfur compared to oxygen. Next, we use the 1$\times$1$\times$2 optimized slab models to investigate how water interacts with each system (Figure \ref{Figure3}).
{\color{black}{Panels (b) and (c) of Figure~\ref{Figure3} depict two different possible configurations for adsorbing water molecules at the interface of the slab with water, either by coordination to the open-metal site (panel b) or by HB formation with organic linkers (panel c). In each panel, the left figure demonstrates the structural changes in Cu$_3$(HHTP)$_2$ due to water adsorption while the right one depicts the corresponding data for Cu$_3$(HTTP)$_2$.  Consistent with our MD simulations, coordination of water to the open--Cu site in Cu$_3$(HHTP)$_2$ is more favorable and is exothermic (as much as 1.7 kcal/mol) than Cu$_3$(HTTP)$_2$. As expected, adsorption of water on top of the slab at the interface does not change the interlayer distance considerably. Panel (c) of Figure~\ref{Figure3} reveals that adsorption of water by HB formation does not solely occur by the linkers of the outermost layer of the slab. But, the organic linkers of the immediate layer beneath are also  sharing the water molecule with the linkers of the outermost layer in both MOFs, Figure \ref{Figure3}. Comparison between panels c and b shows that this double HB isomer is 2.5 and 1.2 kcal/mol more stable than the coordinated isomer in Cu$_3$(HHTP)$_2$ and Cu$_3$(HTTP)$_2$, respectively. This further confirms our classical MD simulation data that HB formation to the organic linkers is the major form of water adsorption (Figure \ref{Figure2}). We also considered the case where the confined water is penetrated within the interlayer space and is simultaneously coordinated to the open Cu site beneath the outermost layer, panels (d) in Figure \ref{Figure3}. The penetrated water molecule in the interlayer space is shared between two open--Cu sites which in contrast to two previous isomers is thermodynamically endothermic. The interlayer distance is expanded around 1.4 \AA~in the area around the coordination site  in both MOFs to accommodate the incoming water molecule. This leads to weakening of the stabilizing $\pi-\pi$ interactions. Increasing the number of water molecules to one per Cu$^{2+}$ center in one layer, \textit{i.e.} 3 water molecules, enhances the structural deformation in both MOFs. Now, the interlayer space between two outermost layers is increased across the unit cell, panel (e) in Figure \ref{Figure2}. Comparing the results from panel (d) and panel (e) it does not seem probable that water is able to completely saturate the interlayer space and separate the two outermost layers. The PBE-D3 calculated $\Delta E_{ads}$s are 10.5 and 18.4 kcal/mol more endothermic than one water adsorption in Cu$_3$(HHTP)$_2$ and Cu$_3$(HTTP)$_2$, respectively. On the other hand, the higher $\Delta E_{ads}$ of Cu$_3$(HTTP)$_2$ than Cu$_3$(HHTP)$_2$, +26.8 vs. +10.9 kcal/mol, confirms the less affinity of the former for interacting with water molecules illustrating its higher stability in aqueous solutions.}}

{\color{black}{\subsection{Electrical conductivity characterization in presence and absence of water}

\begin{figure*}[!h]
\centering
  \includegraphics[width=0.99\linewidth]{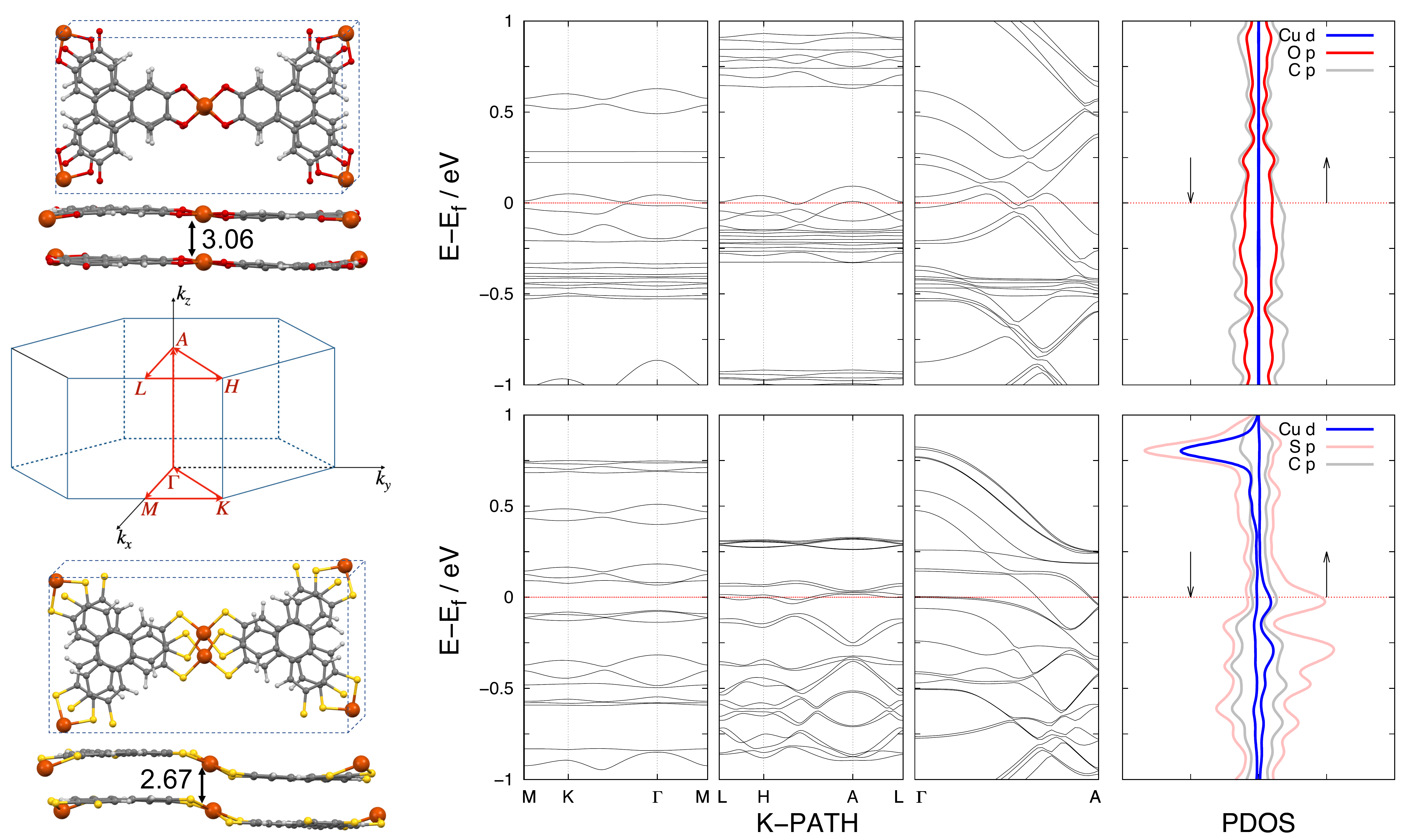}
  \caption{{\color{black}{Band structures and projected density of states (PBE+U) of the dry 1$\times$1$\times$1 Cu$_3$(HHTP)$_2$ (top) and Cu$_3$(HTTP)$_2$ (bottom) 2D MOFs. The Fermi level is highlighted with a dashed red line with key distances (in ~\AA) given.}}}
  \label{fig:dd}
\end{figure*}

Next, we investigate the effects of water adsorption on electrical conductivity of these materials for applications in aqueous solutions, e.g. as electrodes or electrochemical sensors.
In layered 2D materials, the atoms forming the compound are arranged into planes (layers) that are held together by strong in-plane bonds, usually covalent. 
To form a 3D crystal, the layers are stacked in the out-of-plane direction with weak van der Waals interactions.
As a consequence, two different descriptions are usually used for the hopping in a 2D layered semiconductor: through-bond and through-space. The former promotes charge transport 
via favorable spatial and energetic overlap of the metal and ligand orbitals involved in covalent bonding. 
The latter enables charge transport via non-covalent interactions (such as $\pi-\pi$ stacking). 

Dealing with 2D MOFs, the peculiar architecture of the system
provides both \textit{in}-plane and \textit{out}-of-plane electrical conductivity. 
More specifically, as reported by Chen, $et~ al.$~\cite{Chen:2015} they exhibit typical $\pi$-conjugated characteristics, since their projected density of states (pDOS) near the 
Fermi level are almost fully contributed by the $p_z$ orbitals of the ligand as well as the delocalized $d$ orbitals of the transition metal atoms.
Yet, this aspect should be contrasted with the role played by metal hybridization. Indeed, it was previously~\cite{Chen:2015} shown that Ni$^{2+}$ ($d^9$) atoms in the Ni$_3$(HITP)$_2$ sheet adopt the $dsp^2$ hybridization to form a square-planar geometry with the organic moieties which allows perfect 2D $\pi-$ conjugation. 
Conversely, Cu$^{2+}$ (d$^9$) in Cu$_3$(HITP)$_2$ sheet adopts a $sp^3$ hybridization which results in formation of a rather distorted 2D sheet.
\begin{figure*}[!h]
\centering
  \includegraphics[width=0.99\linewidth]{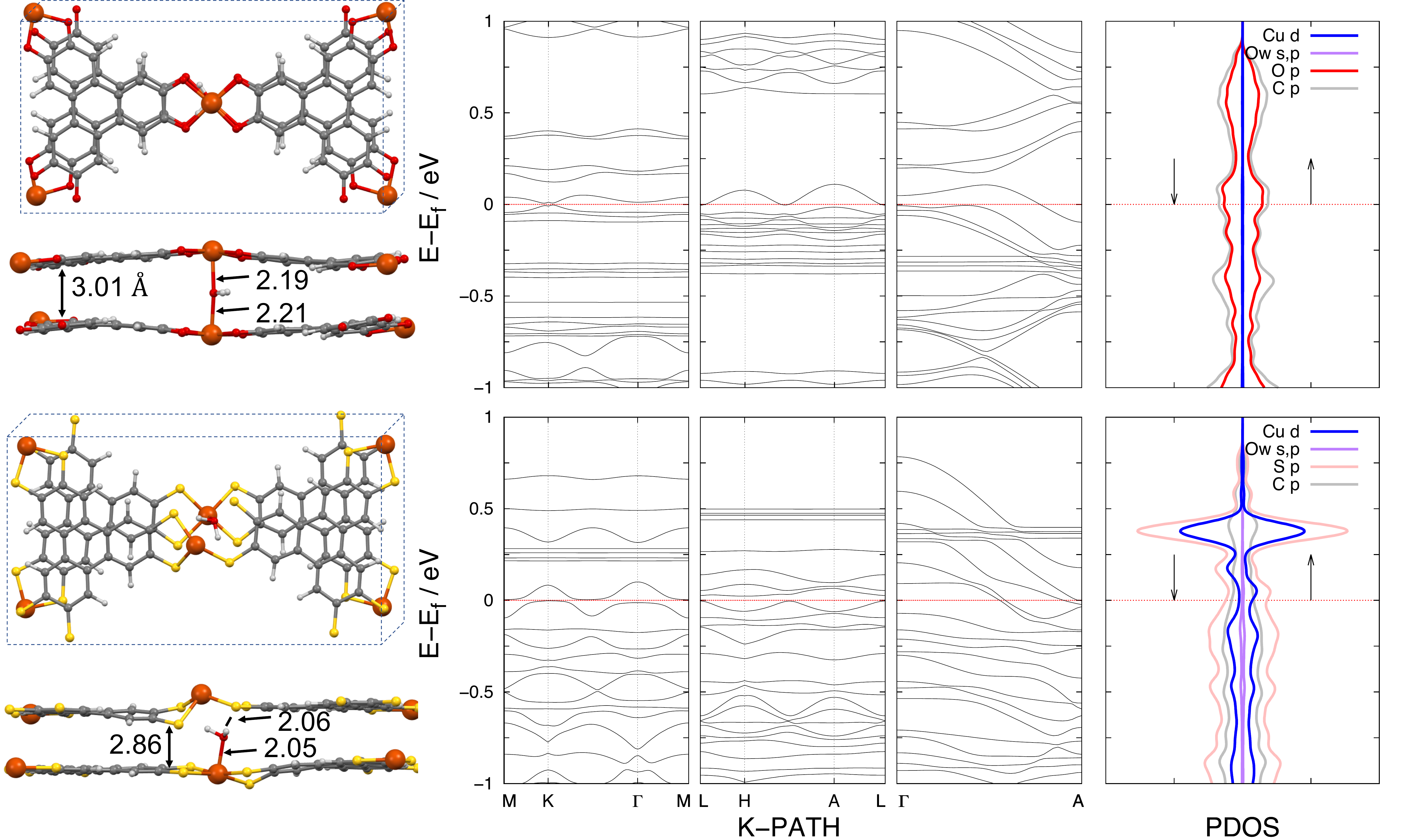}
  \caption{{\color{black}{Band structures and projected density of states (PBE+U) of the mono-hydrated 1$\times$1$\times$1 Cu$_3$(HHTP)$_2$ (top) and Cu$_3$(HTTP)$_2$ (bottom) 2D MOFs. The Fermi level is highlighted with a dashed red line with key distances (in ~\AA) given.}}}
  \label{fig:d1w}
\end{figure*}

The question is how to discern the intra/inter layer contributions in order to account for the intra/inter layer anisotropy in 2D MOFs.\cite{wang19}
 In intrinsic semiconduction, both thermally and optically excited electrons contribute to the conduction. In the absence of photonic excitation, intrinsic semiconduction takes place at temperatures above 0 K as sufficient thermal activation is required to transfer electrons from the valence band to the conduction band.
Then, one may think to monitor the bands crossing at the Fermi level to distinguish  the out-of-plane from the in-plane contributions.
An alternative way consists of the extrinsic investigation, in which the insertion of  impurities in the system
may result in deformations (or their lifts) in the intra- and interlayer regions and provide with useful insights dealing with charge transport.
In addition, the guest molecules can act as charge carriers themselves, as in the case of ionically conductive MOFs, even if the use of guest molecules in a given material 
inevitably reduces the porosity.~\cite{Sun:2016}

From a structural point of view, the two systems under investigation differ in two main aspects (see Figure~\ref{fig:dd}): 
(a) along the stacking axis, the slight dihedral torsion around the central Cu atom in Cu$_3$(HHTP)$_2$  with respect to the main molecular axis,
which must be compared to the dihedral torsion $plus$ deformation dealing with Cu$_3$(HTTP)$_2$; 
(b) along the main molecular axis, the second layer of Cu$_3$(HTTP)$_2$ that slides with respect to the first; by contrast in Cu$_3$(HHTP)$_2$ the two layers are eclipsed.
Another distinction deals with the ligand nature. Indeed, previous works have shown for instance that the active site for catalytic O$_2$ reduction in Ni$_3$(HITP)$_2$ is not metal-based, 
as proposed for many transition-metal macrocycles, but rather ligand-based.~\cite{Miner:2017}

Figure~\ref{fig:dd} displays band diagrams as well as pDOS plots for dry  1$\times$1$\times$1 
unit cells of the Cu$_3$(HHTP)$_2$ and Cu$_3$(HTTP)$_2$ 2D MOFs. 
In intralayer regions, the comparison between the band gap along the MK$\Gamma$ line for Cu$_3$(HTTP)$_2$ and the noteworthy semimetallic signature of Cu$_3$(HHTP)$_2$
suggests a way to modify the conductive nature of MOFs through the ligand. The overlap between deep valence bands (VBs) (the same goes for  conduction bands (CBs)) in MK$\Gamma$ region of
Cu$_3$(HTTP)$_2$ as well as the two valleys on each side of the band gap point toward the hopping options along this line.
But from the curvature of CB and VB we can infer that along the main molecular axis the mobility is higher for Cu$_3$(HHTP)$_2$ than Cu$_3$(HTTP)$_2$. $\pi-\pi$ interactions along the $z$ axis do not look affected by slipping deformations and account for relative larger dispersions in the interlayer region.

The pDOS and band diagrams of hydrated 1$\times$1$\times$1 
unit cells of Cu$_3$(HHTP)$_2$ and Cu$_3$(HTTP)$_2$ are shown in Figure~\ref{fig:d1w}.
They both  disclose a dramatic change in the mobility inserting one water molecule.
The impact on mobility is three fold: (a) the interlayer dispersion is  reduced, especially in the case of Cu$_3$(HTTP)$_2$, in agreement with previous studies~\cite{Foster:2016} which shows that the increase of interlayer distance reduces the interlayer hopping and likely affects the mobility in 2D MOFs.
(b) Intralayer band gaps open up at the Fermi level for both systems along the LHA line, i.e. the MOF's conductive behaviour deviates from (semi)metallic to semiconductor in this region; 
instead, in MK$\Gamma$ region of Cu$_3$(HTTP)$_2$ the gap shrinks moving from dry to hydrated structure. The common feature is the presence of $in-$plane indirect gaps, which means that the conservation of the momentum becomes phonon assisted in this region. 
The different electron/hole mobilities at the band gap reflect a difference in the electron/hole transition 
time and then a decrease in the recombination rate, which in turn favours the conductivity. 
In addition, the  valleys in the conduction band suggest charge carrier traps~\cite{haneef20} likely triggered by the 
guest molecule,  resulting in photogating enhancement and improved (local) photoconductance.~\cite{masur15}
(c) Decreasing the ligand affinity, the $in-$plane mobility at the band gap is increased inserting one water molecule, as we can infer from the band curvatures.
In the case of Cu$_3$(HTTP)$_2$,  the presence of the water molecule reduces the contributions of $p-$ orbitals at the band gap. As mentioned in (b), band structures add significant insights on how to make the most out of guest molecules 
by enhancing a conductive behaviour in some specific regions.
Finally, deformations as well as interlayer displacements seem to affect only through-bond transfers, 
and this provide us with a very useful information on how to tailor band gap openings at the Fermi level.}}

\section{Concluding Remarks}
Overall, using a combination of extensive classical molecular dynamics simulations and periodic quantum mechanics calculations the structure, dynamics and thermodynamics 
of the interfaced vs. confined water in bulk and slab models of 2D MOFs was carefully studied. 
Theory predicts that water is preferentially adsorbed on the framework walls via hydrogen bond formations with the organic linkers rather than coordination 
to the coordinatively unsaturated open-Cu$^{2+}$ sites. Our results show that the interlayer van der Waals interactions are stronger in Cu$_3$(HTTP)$_2$ compared to its Cu$_3$(HHTP)$_2$ 
analogue which helps the MOF to maintain its layered morphology with allowing very little water molecules to diffuse into the interlayer space. 
These information on the properties of the interfaced vs. confined water may be employed to taylor 2D MOFs for specific applications, such as catalytic reactions in operando conditions.

Beyond the influence of water molecules on charge mobility in  MOFs, our findings also highlight the potential of both ligands and guest molecules 
as through-bond descriptors able to discern intra/inter layer anisotropies in 2D MOFs and suggest strategies for designing semiconducting materials.
Nonetheless, assessments concerning electrical conductivity require caution, as the latter is influenced by at least four factors, largely in competition with each other.
(a) A large energy gap, even if detrimental to light absorption, permits efficient charge separation. Electrons (holes) can 
follow the conduction (valence) band profile and possibly reach a charge carrier layer, unless (b) they encounter on their way phonons or impurities 
which in turn (c) may generate charge transport traps that enhance the (local) conductance (i.e., decreasing the resistance of the material).
This does not necessarily translate into an increase of the current, as the latter depends on (d) the recombination rate between valleys on the opposite sides of the 
band gap. For instance, in hydrated Cu$_3$(HTTP)$_2$ the opposite valleys exhibit different curvatures and then different mobilities, 
so one may reasonably expect a decrease in the recombination rate 
which is beneficial to the charge transport. Yet the band gap shrinks after insertion of one water molecule so it is difficult to argue which 
of the two factors may prevail. Therefore, the discussion of this intricate issue would deserve, a thorough quantitative analysis through calculations of effective masses and mobilities, a formalism able to take these four factors into account on equal footing.
Suggestions for improving charge transport in 2D MOFs may rely on purely chemical intuition, for instance by increasing the molecular overlaps, i.e. optimizing the crystal packing (through pressure or strain).
Further thickness reduction to a small number of layers induces quantum confinement in the vertical direction, 
which has a strong impact on the bandgap of the material. 
Or by increasing the intermolecular vibration frequency by tightening the intermolecular bonds or multi-valley schemes nearby the Fermi level.
Studies in these directions are currently underway in our group.

\section{Computational Methods}
A summary of our theoretical methodology is provided below with more details given in the Supporting Information (SI).

\subsection{Classical molecular dynamics simulations} 
Classical molecular dynamics (MD) simulations for dry MOFs were initiated from the hexagonal ($\alpha=\beta=90\degree$, $\gamma=120\degree$, $a = b = 46.470$ \AA, and $c = 13.555$ \AA) PBE-D3 minimized tetra-layered 2$\times$2$\times$2 supercell of the bulk Cu$_3$(HHTP)$_2$ and Cu$_3$(HTTP)$_2$ MOFs (comprised of 1008 atoms in total and 48 metal centers) using DL\_POLY\_2 package.\cite{Smith:1996} {\color{black}{Force field parameters for transition metals involved in SBUs of our 2D MOFs do not exist in Generic force fields such as generalized amber force field (GAFF).\cite{Wang:2004} The first step for generating these parameters is to create a reliable training set of SBUs at \textit{ab initio} level.}} Starting from the only experimentally available crystallographic data,  for Co$_3$(HHTP)$_2$,\cite{Hmadeh:2012} we built a 2$\times$2$\times$2 super cell for the dry MOF by removing hydrolyzed layers, and all chemisorbed water molecules, replacing Co atoms with Cu and adjusting the interlayer distances to $\sim$4 \AA. Both cell vectors and atomic positions of this supercell were then minimized with periodic boundary conditions using the Perdew–Burke–Ernzenhof (PBE)\cite{PBE} density functional with damped D3 dispersion correction\cite{DFTD3} in CP2K version 5.1.\cite{Hutter:2014}. For creating the training sets, reduced cluster models comprised of a single Cu center and two truncated linkers were cut from the optimized crystal structures (see SI Figure S1). The central metal atoms were then displaced by 0.02 \AA~from -0.04 \AA ~to +0.04 \AA ~along the \textit{x}, \textit{y} and \textit{z} dimensions (creating a total of 125 configurations). All electronic energies for this training set were calculated using the $\omega$B97M-v\cite{Mardirossian:2016} density functional and the def2-TZVP basis set as implemented in QCHEM version 5.2.\cite{Shao:2015} To fit force field parameters to this \textit{ab initio} training set, Morse potential was used for all coordinative Cu--O and Cu--S bonds while Harmonic potential was employed for the rest. David Carrol’s genetic algorithm\cite{Goldberg:1989} was used for fitting all bonded interactions involving the Cu$^{2+}$ transition metal centers, including bonds, angles, and dihedrals, while all parameters related to the intramolecular interactions present in the organic linkers were taken from GAFF without further modification. The non--bonded parameters concerning transition metal sites include electrostatic as well as van der Waals interactions. To account for the former, we computed atomic charges at the $\omega$B97X-d/def2-TZVP level\cite{Mardirossian:2016} using the CHELPG scheme which fits all atomic charges to represent molecular electrostatic potential.\cite{Breneman:1990} For the latter, Lennard-Jones parameters of Cu$^{2+}$ were taken from the Universal Force Field (UFF).\cite{Rappe:1992} More details are given in the SI section S1 as well as validation of the developed force fields are provided in SI section S2. The complete list of the bonded and non--bonded parameters for Cu$_3$(HHTP)$_2$ and Cu$_3$(HTTP)$_2$ MOFs are given in SI Tables S1-S5 and Tables S6-S10, respectively.
A second 2D slab model comprised of eight layers and 2016 atoms was built for both systems to compare the properties of slab and bulk in water adsorption and dissolution. A vacuum space of 40 \AA~was added to the \textit{c} vector of the slab models. Each system was equilibrated for 5 ns in the isothermal--isobaric NPT ensemble with a time step of 0.2 fs at 293 K temperature and 1 atm pressure, the experimental conditions at which the original 2D MOFs were synthesized,\cite{Hmadeh:2012} allowing the simulation box to vary. The constant total energy of the systems, with less than 1 kcal/mol fluctuation confirms that 5 ns is indeed enough in order to reach an equilibrated stage. In each 1D channel of our 2$\times$2$\times$2 supercells, there are $n=24$ open metal sites that are exposed to confined water molecules. Therefore, to study dynamics of confined water in the bulk models, we placed $2n=48$, $4n=96$ and $8n=192$ water molecules in a sphere centered at the middle of the 1D channel of the dry MOF using PACKMOL\cite{Martnez:2009} code. For the slab models, a layer of water was placed in the vacuum above the slab $\sim$5 \AA~ away from the surface. The equations of motion were propagated according to the velocity--Verlet algorithm. The temperature was kept constant using a Nosé--Hoover chain comprised of four thermostats.\cite{Tuckerman:2010} To model water molecules, we use the flexible 4-site qTIP4P/F\cite{Habershon:2009} quantum water potential which has been shown to be successful in reproducing a diverse number of static and dynamical properties of water including melting point, diffusion coefficients and IR spectrum. Dynamical properties are calculated from the average of 10 independent 50 ps NVE simulations. These trajectories were run from 10 different initial configurations obtained from 10 ps NVT trajectories that followed the NPT simulations. The final snapshots of these NVT trajectories were then used as initial configurations for 50 ps NVE simulations ensuring that different starting configurations initiate independent NVE trajectories.

\subsection{Electronic structure calculations} 
The PBE\cite{PBE} exchange-correlation functional corrected by the DFT-D3 method of Grimme\cite{DFTD3} as implemented in Vienna Ab Initio Simulation Package (VASP)\cite{Kresse1993, Kresse1994, Kresse1996, Kresse19962} was used to calculate the band structure and density of states. Hubbard U corrections where U parameter is set to 10.4 eV, suggested by Gregory et al\cite{Gregory:2016}, are included to treat 3d states of the Cu transition metals\cite{Anisimov:1991, Anisimov:1997, Himmetoglu:2014}. Interactions between electrons and ions were described by Projector Augmented Wave (PAW) potentials\cite{PAW1, PAW2} with energy cutoff set to 500 eV. Gaussian smearing was adopted in all calculations with a smearing width of 0.05 eV. The convergence criteria were 10$^{-5}$ for self consistent field calculations and $10^{-6}$ for electronic property calculations. A \textit{k}-point mesh in the Monkhorst--Pack scheme of 2$\times$2$\times$6 was used in the SCF part and twice denser in following calculations. Spin polarized calculations (collinear) were performed for all systems.

\begin{suppinfo}
Details of our classical MD simulations and quantum mechanical calculations. 
\end{suppinfo}

\begin{acknowledgement}
This work is supported by start-up fund from NJIT and used the Extreme Science and Engineering Discovery Environment (XSEDE) which is supported by NSF grant numbers CHE200007 and CHE200008.  This research has been (partially) enabled by the use of computing resources provided by Kong HPC center at NJIT. 
\end{acknowledgement}

\bibliography{rsc}
\bibliographystyle{achemso}
\clearpage

\end{document}